\documentstyle[epsfig]{mn}

\newif\ifAMStwofonts

\ifoldfss
  \newcommand{\rmn}[1] {{\rm #1}}

  \ifCUPmtlplainloaded \else
    \NewTextAlphabet{textbfit} {cmbxti10} {}
    \NewTextAlphabet{textbfss} {cmssbx10} {}
    \NewMathAlphabet{mathbfit} {cmbxti10} {} 
    \NewMathAlphabet{mathbfss} {cmssbx10} {} 
  \fi
  \ifAMStwofonts
    \ifCUPmtlplainloaded \else
      \NewSymbolFont{upmath} {eurm10}
      \NewSymbolFont{AMSa} {msam10}
      \NewMathSymbol{\upi}     {0}{upmath}{19}
      \NewMathSymbol{\umu}     {0}{upmath}{16}
      \NewMathSymbol{\upartial}{0}{upmath}{40}
      \NewMathSymbol{\leqslant}{3}{AMSa}{36}
      \NewMathSymbol{\geqslant}{3}{AMSa}{3E}

    \fi
  \fi
\fi 

\ifnfssone
  \newmathalphabet{\mathit}
  \addtoversion{normal}{\mathit}{cmr}{m}{it}
  \addtoversion{bold}{\mathit}{cmr}{bx}{it}
  \newcommand{\rmn}[1] {\mathrm{#1}}

  \newmathalphabet{\mathbfit} 
  \addtoversion{normal}{\mathbfit}{cmr}{bx}{it}
  \addtoversion{bold}{\mathbfit}{cmr}{bx}{it}
  \newmathalphabet{\mathbfss} 
  \addtoversion{normal}{\mathbfss}{cmss}{bx}{n}
  \addtoversion{bold}{\mathbfss}{cmss}{bx}{n}
  \ifAMStwofonts
    \ifCUPmtlplainloaded \else
      \UseAMStwoboldmath
      \makeatletter
      \new@mathgroup\upmath@group
      \define@mathgroup\mv@normal\upmath@group{eur}{m}{n}
      \define@mathgroup\mv@bold\upmath@group{eur}{b}{n}
      \edef\UPM{\hexnumber\upmath@group}
      \new@mathgroup\amsa@group
      \define@mathgroup\mv@normal\amsa@group{msa}{m}{n}
      \define@mathgroup\mv@bold\amsa@group{msa}{m}{n}
      \edef\AMSa{\hexnumber\amsa@group}
      \makeatother
      \mathchardef\upi="0\UPM19
      \mathchardef\umu="0\UPM16
      \mathchardef\upartial="0\UPM40
      \mathchardef\leqslant="3\AMSa36
      \mathchardef\geqslant="3\AMSa3E
    \fi
  \fi
\fi 

\ifnfsstwo
  \newcommand{\rmn}[1] {\mathrm{#1}}

  \DeclareMathAlphabet{\mathbfit}{OT1}{cmr}{bx}{it}
  \SetMathAlphabet\mathbfit{bold}{OT1}{cmr}{bx}{it}
  \DeclareMathAlphabet{\mathbfss}{OT1}{cmss}{bx}{n}
  \SetMathAlphabet\mathbfss{bold}{OT1}{cmss}{bx}{n}
  \ifAMStwofonts
    \ifCUPmtlplainloaded \else
      \DeclareSymbolFont{UPM}{U}{eur}{m}{n}
      \SetSymbolFont{UPM}{bold}{U}{eur}{b}{n}
      \DeclareSymbolFont{AMSa}{U}{msa}{m}{n}
      \DeclareMathSymbol{\upi}{0}{UPM}{"19}
      \DeclareMathSymbol{\umu}{0}{UPM}{"16}
      \DeclareMathSymbol{\upartial}{0}{UPM}{"40}
      \DeclareMathSymbol{\leqslant}{3}{AMSa}{"36}
      \DeclareMathSymbol{\geqslant}{3}{AMSa}{"3E}
    \fi
  \fi
\fi 

\ifCUPmtlplainloaded \else
  \ifAMStwofonts \else 
    \def\upi{\pi}
    \def\umu{\mu}
    \def\upartial{\partial}
  \fi
\fi

\title{The cooling of shock-compressed primordial gas}
\author[J.L. Johnson and V. Bromm]
       {Jarrett L. Johnson\thanks{E-mail: jljohnson@astro.as.utexas.edu} and Volker Bromm \\
 Department of Astronomy, University of Texas, Austin, TX 78712, USA \\}

\pubyear{2005}

\begin{document}

\maketitle
\topmargin-1cm

\label{firstpage}

\begin{abstract}
We find that at redshifts $z \ga 10$, HD line cooling allows strongly-shocked primordial gas to cool to the temperature of the cosmic microwave background (CMB).
This temperature is the minimum value attainable via radiative cooling.
Provided that the abundance of HD, normalized to the total number density,
exceeds a critical level of $\sim 10^{-8}$, the CMB temperature floor
is reached in a time which is short compared to the Hubble time.
We estimate the characteristic masses of stars formed out of shocked primordial gas in the wake of the first supernovae, and resulting from the virialization of dark matter haloes during hierarchical structure formation to be $\sim 10{\rmn M}_{\odot}$.
In addition, we show that cooling by HD enables the primordial gas in relic
H~II regions to cool to temperatures considerably lower than those reached via H$_2$ cooling alone. 
We confirm that HD cooling is unimportant in cases where the
primordial gas does not go through an ionized phase, as in the formation
process of the very first stars in $z\ga 20$ minihaloes of mass $\sim
10^{6}{\rmn M}_{\odot}$.
\end{abstract}


\section{Introduction}
 
What were the properties of the first generations of stars that
formed at the end of the cosmic dark ages?
Within the standard picture of hierarchical structure formation, shocks in the primordial gas arise naturally, both through the merging of dark matter (DM) haloes in the formation of galaxies and through the supernovae (SNe) that marked the deaths of the first generation stars (e.g. Barkana \& Loeb 2001; Ciardi \& Ferrara 2005). 
During the merging of DM haloes, the primordial gas embedded in them can be shock-heated to high temperatures ($\ga 10^{4}$~K), owing to the large velocities with which the haloes collide.  
  Blast waves, triggered by the first SNe in the early Universe, are expected to have been commonplace (e.g. Madau, Ferrara \& Rees 2001; Mori, Ferrara \& Madau 2002; Bromm, Yoshida \& Hernquist 2003), as the first generation of stars were likely very massive (Bromm, Coppi \& Larson 1999, 2002;  Abel, Bryan \& Norman 2002; Nakamura \& Umemura 2001).  
The first stars, called Population~III (Pop~III) stars because they are 
assumed to have near-zero metallicity (e.g. Bromm \& Larson 2004; Glover 2005), are thus
  predicted to be sufficiently shortlived ($\la 3$~Myr) to explode within a time short compared to the Hubble time (e.g. Bond, Arnett \& Carr 1984).  

  It has long been realized that the evolution of shocked primordial gas is markedly different from the nearly adiabatic one in the early minihaloes 
with total masses $\sim 10^{6}{\rmn M}_{\odot}$ and collapse redshifts of $z\ga 20$ inside
  of which the very first stars are predicted to have formed (e.g. Mac Low \& Shull 1986; Shapiro \& Kang 1987; Yamada \& Nishi 1998; Oh \& Haiman 2002; Clarke \& Bromm 2003; Omukai \& Yoshii 2003).  In the strong-shock limit, the shocked, ionized gas cools roughly isobarically, and can cool faster than it recombines.  Under these conditions, molecular hydrogen is readily formed, and the resultant high abundance of H$_2$ allows the gas to cool to $\la 200$ K and to achieve high densities (e.g. Kang \& Shapiro 1992).  The cooling of primordial gas behind shock fronts thus allows gas fragmentation to occur at lower temperatures, possibly leading 
  to the formation of stars with masses that are smaller than in the unshocked case (e.g. Mackey, Bromm \& Hernquist 2003;
  Salvaterra, Ferrara \& Schneider 2004; Machida et al. 2005).

\begin{table*}

\begin{tabular}{ccccc}
\hline
Population  & Z/Z$_{\odot}$ & M$_{\rmn char}$/M$_{\odot}$ & Site of Formation  \\
\hline
III  & 0  & $\sim 100$ & $\sim 10^{6}{\rmn M}_{\odot}$ haloes\\
II.5 & very small or zero & $\sim 10$ & $\ga 10^{8}{\rmn M}_{\odot}$ haloes\\
II   & $\sim$ 10$^{-2}$  & $\sim$ 1 & Early galaxies\\
I    & $\ga 0.1$ & $\sim$ 1 & Present-day galaxies\\
\hline
\end{tabular}
\caption{Stellar populations in the high-redshift Universe. For each population, we show
the typical metallicity of the parent gas, the estimated fragmentation mass scale, and
the typical formation site (see text for a detailed discussion).}
\label{tab1}
\end{table*}

The presence of deuterium in the primodial gas has recently received renewed attention as a means to allow the gas to cool to temperatures yet lower than those possible solely via H$_2$ line cooling (e.g. Flower et al. 2000; Galli \& Palla 2002; Nakamura \& Umemura 2002; 
see also Suchkov, Shchekinov \& Edelman 1983).  
The role of HD in star formation, through the fragmentation of SN-shocked primordial gas, has been studied as well, suggesting that line cooling by HD may facilitate the formation of low mass metal-free stars, possibly even of primordial brown dwarfs (Uehara \& Inutsuka 2000; Machida et al. 2005).  

Employing the extensive deuterium chemical network compiled by Nakamura \& Umemura (2002), we investigate the importance of HD in the cooling of primordial gas, focusing on four distinct cases: shocks from SNe, shocks associated with the virialization of DM haloes in large-scale structure formation, collapse within minihaloes, and collapse within relic H II regions.  In addition, we provide estimates for the characteristic mass of stars formed by the fragmentation of strongly-shocked primordial gas. The outline of our paper is as follows. In Section 2 we discuss the basic physical ingredients included in our treatment of the cooling of shocked primordial gas.  Section 3 describes the evolution of the primordial gas in structure formation shocks and in SN-induced shocks, as well as in minihaloes and relic H II regions.  In Section 4 we derive the characteristic mass of stars formed through the fragmentation of shocked primordial gas.  Finally, we present our conclusions in Section 5.

\section{Basic physical ingredients}
\subsection {Shocks in the high-z Universe}
We consider two different situations in the early universe where the primordial gas is strongly shocked: the assembly of DM-dominated haloes in large-scale structure formation and SN explosions of the first stars.  In each of these cases the primordial gas is shock-heated to high temperatures and ionized, after which it cools roughly isobarically (Shapiro \& Kang 1987; Yamada \& Nishi 1998).  This condition allows the gas to become very dense as it cools behind the shock front, as well as allowing it to cool more quickly than it can recombine.  Thus, the gas remains out of equilibrium even after cooling to below $\sim 10^4$ K, and the formation of molecular hydrogen is enhanced, catalyzed by free electrons and hydrogen ions (see Kang \& Shapiro 1992).
The result is an enhanced fraction of H$_2$, the line cooling of which can then efficiently lower the gas temperature all the way down to $\la 200$ K.  Without sufficient H$_2$, gas of primordial composition can obtain, via atomic hydrogen 
line cooling, temperatures no lower than $\sim 10^{3.9}$ K.

The high densities and low temperatures attainable by the shocked gas in this way, through isobaric cooling and the effective out-of-equilibrium formation of molecular hydrogen, make possible a much lower Jeans mass than would otherwise be the case for a primordial gas cloud, since $M_{\rm J}\propto T^{3/2}n^{-1/2}$, where $T$ is the temperature of the gas and $n$ is the number density of its constituent particles.  This, in turn, translates into lower characteristic masses for the stars that form from such shocked primordial gas.  

This is one mechanism that would lead to the formation of so-called Population II.5 (Pop~II.5) stars, which have been postulated to have formed from the primordial gas shocked by the SN explosions of Pop~III stars (Mackey et al. 2003; Salvaterra et al. 2004).  More generally, Pop~II.5 stars could have formed from primordial gas that has been strongly ionized, thus allowing for the free electron-catalyzed formation of molecules and so for enhanced cooling of the primordial gas, regardless of the source of the ionization.  Another possible site for the formation of Pop~II.5 stars thus are DM haloes of masses $\ga 10^8$ M$_{\odot}$, containing gas shocked through the hierarchical merging of DM haloes in the process of early structure formation, at redshifts $z \ga 10$.  Finally, Pop~II.5 stars may also have formed in the relic H II regions of the first stars, as the formation of molecules would have been catalyzed by the residual fraction of free electrons in the cooling gas following the death of the central massive star (e.g. O'Shea et al. 2005).  
Pop~II.5 stars would have been of almost purely primordial composition and with masses at least an order of magnitude lower than the typical masses of Pop~III stars.  A careful treatment of the primordial chemistry, including the cooling effects of HD, appears to be necessary in determining the characteristic masses of Pop~II.5 stars.

To summarize, the postulated Pop~II.5 stars, defined as stars formed from strongly ionized primordial gas, have masses intermediate between
Pop~III and Pop~II ones, but may still be almost metal-free.
The distinct modes of star formation are discussed in Table~1.

\subsection {Chemistry of the primordial gas}
In evaluating the cooling properties of shocked primordial gas, we adapt the one-zone model used in Mackey et al. (2003).  This model solves for the thermal and chemical evolution of isobarically cooling gas of primordial composition, though insofar only as including normal hydrogen and helium (see Mackey et al. 2003; Bromm et al. 2002).  In addition to these chemical species, we have included for our calculations the five deuterium species D, D$^{+}$, D$^{-}$, HD, and HD$^{+}$, together with a network of 18 reactions in which these species participate.  The rate coefficients for these reactions are given in table 1 of Nakamura \& Umemura (2002). 
We replace the rate coefficients for the following reactions with the updated rate coefficients from Galli \& Palla (2002):  

\begin{displaymath}
{\rmn D} + {\rmn H^{+}} \to {\rmn D^{+}} + {\rmn H} \mbox{\ ,}
\end{displaymath} 

\begin{displaymath}
{\rmn D^{+}} + {\rmn H} \to {\rmn D} + {\rmn H^{+}} \mbox{\ ,}
\end{displaymath} 

\begin{displaymath}
{\rmn D^{+}} + {\rmn H_{2}} \to {\rmn H^{+}} + {\rmn HD} \mbox{\ ,}
\end{displaymath} 

\begin{displaymath}
{\rmn HD} + {\rmn H^{+}} \to {\rmn H_{2}} + {\rmn D^{+}} \mbox{\ .}
\end{displaymath} 

We would like to point out that the rate coefficient for reaction (3), as quoted in Galli \& Palla (2002), contains a typo, and we refer the reader to the original reference for this rate (Savin 2002) for the correct value. 

\begin{figure}
\vspace{2pt}
\epsfig{file=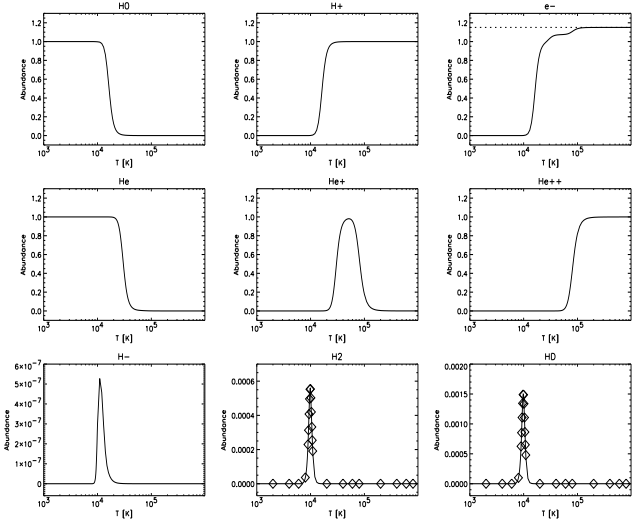,width=8.5cm,height=7.cm}
\caption{The equilibrium abundances of the major species included in calculating the evolution of primordial gas.  The solid lines show the equilibrium abundances to which our chemical network converged, while the diamond points are the results from analytical calculations of the equilibrium abundances of the two most important coolants of the gas at low temperatures, H$_2$ and HD.  The electron abundance exceeds unity at high temperatures because of the contribution of electrons due to the ionization of helium at those high temperatures.  The dashed line in the panel for the electron abundance denotes the maximum abundance possible for the electrons, which occurs only when helium is completely ionized.    
}
\end{figure}
We expect that the most important coolant in the primordial gas at temperatures $\la 100$ K is HD, and hence we neglect the presence of other trace species which could have formed in the primordial gas, such as LiH owing to the very low primordial abundance of lithium ($n_{\rmn Li}/n \sim 10^{-10}$).  We also neglect the cooling of the primordial gas due to ionic molecules such as H$_2$$^+$, H$_3$$^+$, and HeH$^+$, which could be significant in the temperature range 5,000 to 8,000 K. Excluding these coolants
only strengthens our result, since any additional cooling at $<8,000$~K would enhance the out-of-equilibrium
production of H$_2$ and HD.

Our chemical network was tested by verifying that the equilibrium abundances of the main coolant species, as calculated numerically, matched those derived analytically.  The results of these calculations are displayed in Fig.~1, the solid lines showing the equilibrium abundances converged upon by the code and the diamonds representing the analytical ones.  We limit our analytical derivations to the two most important coolants at the low temperatures at which the gas fragments ($ \la 200$ K), H$_2$ and HD.  As Fig.~1 shows, the numerical results nicely reproduce the analytical values (for details of the analytical estimates see Bromm et al. 2002).             

 The primary reaction sequence which creates HD in the shocked primordial gas is (e.g. Galli \& Palla 2002; Machida et al. 2005) 
  
 \begin{figure}
\vspace{2pt}
\epsfig{file=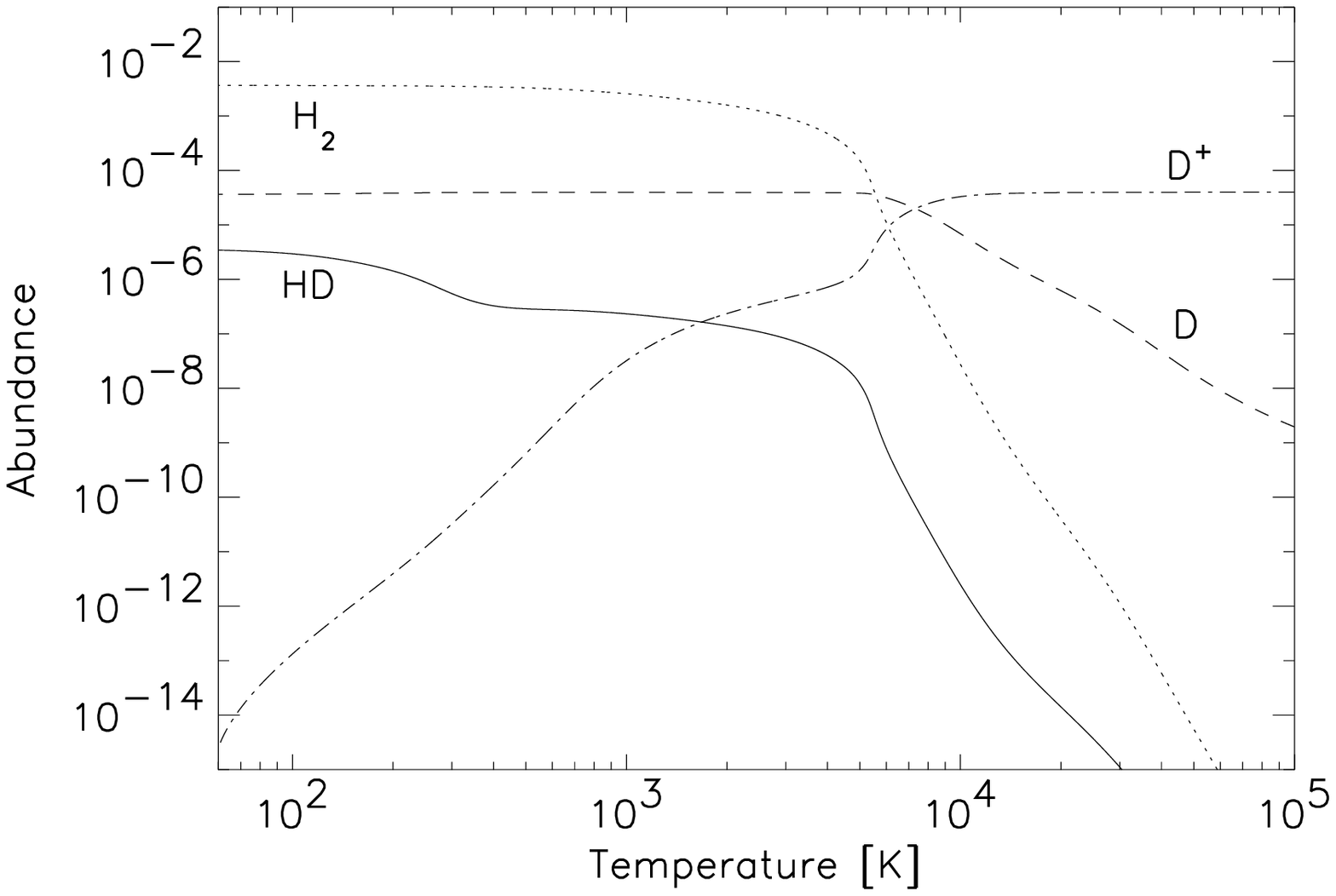,width=8.5cm,height=7.cm}
\caption{The abundances of HD, H$_2$, D, and D$^+$ as a function of temperature, corresponding to the 100 km s$^{-1}$ shock at {$z$ = 20} (see lower left panel of Fig.~3).  The dominant reaction forming HD is {D$^+$ + H$_2$ $\to$ HD + H$^+$}.  Despite the steady decline in the abundance of D$^+$ with decreasing temperature, HD is still created predominantly through the above reaction.}
\end{figure}

\begin{equation}
{\rmn e^{-}} +{\rmn  H} \to {\rmn H^{-}} + h\nu \mbox{\ ,}
\end{equation} 
       
\begin{equation}
{\rmn H^{-}} +{\rmn  H} \to {\rmn H_{2}} + {\rmn e^{-}} \mbox{\ ,}
\end{equation} 

\begin{equation}
{\rmn D^{+}} +{\rmn  H_{2}} \to {\rmn HD} + {\rmn H^{+}} \mbox{\ .}
\end{equation} 
The dependence of the HD abundance on those of H$_2$ and D$^+$ is illustrated in Fig.~2, which shows the abundances of H$_2$, HD, D, and D$^+$ as a function of temperature in the case of a SN-induced shock, corresponding to the lower left panel of Fig.~3. Evident from Fig.~2 is a steady increase in HD abundance with lower temperature, and so with higher density, at temperatures $T_{\rmn gas} \la 500$~K, which is due to the out-of-equilibrium abundance of H$_2$ and the residual abundance of D$^+$ which has not recombined even at the lowest temperatures $T_{\rmn gas} \simeq T_{\rmn CMB}$. 

\subsection{Photodestruction of HD}
It is well established that photons in the Lyman-Werner (LW) bands, with energies 
between 11.2 and 13.6 eV, can easily dissociate H$_2$ molecules (e.g. Haiman, Rees
\& Loeb 1997; 
Bromm \& Larson 2004 and references therein).  Here we evaluate the effect that such radiation has in 
dissociating HD molecules.

Regardless of the source of the radiation, the timescale for the photodissociation of HD by LW photons can be approximated by $t_{\rm diss} \simeq 1000/\chi$~yr, where $\chi$ is the LW radiation field in units of the Draine standard field (see figure~1 in Petit, Roueff \& Le Bourlot 2002). We obtain the value of the Draine standard LW field by integrating the photon flux defined by equation (11) of Draine (1978) from 11.2 eV to 13.6 eV, which yields $\sim 10^{-20}$ erg s$^{-1}$ cm$^{-2}$ Hz$^{-1}$ sr$^{-1}$.  Thus, for a general radiation field the HD photodissociation time can be expressed as $t_{\rm diss}$ $\sim$ 10$^8$ yr ($J_{\rmn LW}$/10$^{-4}$)$^{-1}$, where $J_{\rmn LW}$ measures the LW flux in units of 10$^{-21}$ erg s$^{-1}$ cm$^{-2}$ Hz$^{-1}$ sr$^{-1}$.

To estimate the importance of radiation emitted by shocked primordial gas 
in destroying HD molecules, we compare the HD photodissociation timescale, 
$t_{\rm diss}$, to the formation timescale, $t_{\rm form}$
(see Machida et al. 2005).  The 
LW photon flux emergent from a 100 km s$^{-1}$ shock passing through 
a medium of pre-shock hydrogen number density of $10^2$ cm$^{-3}$ is $\sim
10^2$ photons cm$^{-2}$ s$^{-1}$, which corresponds to $J_{LW} \sim 10^{-4}$ (Shull \& Silk 1979).  This results in 
$t_{\rm diss}$ $\sim$ 10$^8$ yr 
(Le Petit, Roueff \& Le Bourlot 2002). From our calculations, we estimate 
$t_{\rm form} \sim 10^{6}$ yr for the same shock velocity 
and pre-shock density.  Thus, for cases in which the post-shock gas is the 
only source of LW photons, the formation timescale is shorter 
than the dissociation timescale by two orders of magnitude, and the effect of 
photodissociating radiation produced by the shock is likely to be unimportant 
over the range of shock velocities and pre-shock densities that we consider 
in this paper.
  
 We note, however, that in relic H II regions the photodestruction of HD may be particularly important due to emission of LW radiation through the two-photon decay of recombining helium atoms, at densities below $\sim$ 10$^4$ cm$^{-3}$ (Mathis 1957; Pottasch 1961).  While helium recombination radiation will also be generated by shocks strong enough to ionize helium, the column densities of ionized helium in the first relic H II regions are expected to be much higher than those in shock fronts, since the first H II regions generated by massive Pop III stars are expected to have been of order a few kpc in diameter with the majority of the helium being ionized by the hard UV emission from the central star (Kitayama et al. 2004; Whalen, Abel \& Norman 2004; Alvarez, Bromm \& Shapiro 2005). With these larger column densities, the fluxes of recombination radiation from helium in relic H II regions may be strong enough to efficiently photodissociate HD molelcules.  However, if molecule formation takes place efficiently after the gas has cooled below the temperatures at which helium recombines, then the effect of the destruction of molecules by helium recombination radiation could be mitigated.  A careful treatment of the effect of recombination radiation is clearly required in order to accurately calculate the chemical and thermal evolution of relic H II regions. We defer such a treatment to future work.     

If the column density of HD is high enough that self-shielding is important, the value of $t_{\rm diss}$ calculated here will represent a conservative lower limit on the photodissociation time of shielded HD molecules.  The effects of H$_2$ self-shielding have been widely studied and have been found to have important consequences for the cooling of primordial gas subject to a LW background radiation field (Draine \& Bertoldi 1996; Haiman et al. 1997; Yoshida et al. 2003). As well, it has been shown that the dissociation rate of HD shielded by H$_2$ is expected to be lower than the dissociation rate of unshieldded HD by up to a factor of about 1/3, due to the close overlapping of several HD lines by H$_2$ lines (Barsuhn 1977).  Here, for simplicity, we neglect the effects of both HD self-shielding and HD shielding by H$_2$. We note, however, that if there exists a general LW background in addition to the radiation emitted locally by the shocked 
gas, such as from first generation massive stars during structure formation or from continuum radiation from recombining helium atoms in relic H II regions, then self-shielding effects may become much more important in preventing the photodissociation of HD.    

\subsection {Cooling by HD}
The cooling properties of HD are distinct from those of H$_2$ for the following reasons (e.g. Flower et al. 2000; Uehara \& Inutsuka 2000).  Firstly, HD has a permanent dipole moment, allowing dipole rotational transitions, which spontaneously occur much more often than the quadrupole rotational transitions in H$_2$.  Also, the dipole moment of HD allows rotational transitions of $\Delta J = \pm 1$, which are of lower energy than the $\Delta J = \pm 2$ quadrupole transitions of H$_2$.  Thus, collisional excitation of HD from the ground to the first excited rotational level ($J$=1), and the subsequent radiative decay back to the ground state by a dipole transition, can allow HD to efficiently cool gas to lower temperatures than can be reached with H$_2$ line cooling alone.  Finally, the rotational energy levels of HD are lower than those of H$_2$ by a factor of the reduced mass, which is higher by a factor of $\sim 4/3$ for HD than for H$_2$, allowing for even lower energy collisions to excite the first rotational level.  

We have implemented the HD cooling function provided by Flower et al. (2000; see also Lipovka et al. 2005).  To take into account that the gas cannot radiatively cool to below the temperature of the cosmic microwave background (CMB), 

\begin{equation}
T_{\rmn CMB} = \mbox{2.7~K}(1+z)\mbox{\ ,}
\end{equation}
we employ an effective cooling rate, according to      

\begin{equation}
\Lambda = \Lambda (T_{\rmn gas})-\Lambda (T_{\rmn CMB})\mbox{\ ,}
\end{equation}
where $T_{\rmn gas}$ is the temperature of the gas. 

\subsection {The CMB cooling limit}
For a gas which cools purely radiatively, and not by adiabatic expansion, the CMB sets a lower limit on the temperature to which it can cool (e.g. Larson 1998).  To show this, we will consider the process by which the primordial gas cools to temperatures approaching the CMB value. 
The frequency of emitted radiation for the rotational transition $J = 1 \to 0$ of HD is given by 

\begin{equation}
\frac{h\nu_{10}}{k_{\rmn B}} \simeq  \mbox{130 K}\mbox{\ ,}
\end{equation}
where $\nu_{10}$ is the frequency for the $J=1 \to 0$ rotational transition, $k_{\rmn B}$ is the Boltzmann constant, and $h$ is the Planck constant.  We investigate how the primordial gas approaches $T_{\rmn CMB}$ by only considering this transition and its reverse.  This is reasonable since all other transitions from the ground state of HD have excitation energies significantly higher than that for this transition and since, for the redshifts we are considering, $T_{\rmn CMB} < 130$ K.

Denote the Einstein coefficients for spontaneous and stimulated emission from $J=1 \to 0$ by $A_{10}$ and $B_{10}$, respectively.  The Einstein coefficient for absorption of a photon effecting the transition $J=0 \to 1$ is $B_{01}$.  Now, consider a finite parcel of optically thin primordial gas at a temperature $T_{\rmn gas}$. 
We assume for simplicity that densities are sufficiently high
to establish LTE level populations 
according to the Boltzmann distribution

\begin{equation}
\frac{n_{1}}{n_{0}} =  \frac{g_{1}}{g_{0}} e^{-\frac{h \nu_{10}}{k_{\rmn B} T_{\rmn gas}}} \mbox{\ ,}
\end{equation}
where $g_{1}$ and $g_{0}$ are the statistical weights of the first excited and ground rotational levels with $g_{1} = 3 g_{0}$.
Furthermore, it is assumed that only the ground and first excited rotational levels are occupied.  That is, we take it here that already $T_{\rmn gas} < 130$ K. We also make the assumption that $T_{\rmn gas} \ga T_{\rmn CMB}$.  Thus, if we denote the blackbody specific intensity of the CMB at the frequency $\nu_{10}$ as $I_{\nu_{10}}$, then it follows that

\begin{equation}
I_{\nu_{10}} = \frac{2h\nu_{10}^{3}/c^2}{e^{\frac{h\nu_{10}}{k_{\rmn B} T_{\rmn CMB}}}-1} < \frac{2h\nu_{10}^{3}/c^2}{e^{\frac{h\nu_{10}}{k_{\rmn B} T_{\rmn gas}}}-1}\mbox{\ .}
\end{equation}   
If we use the standard relations $B_{10}g_{1} = B_{01}g_{0}$ and 

\begin{equation}
\frac{2h\nu_{10}^{3}}{c^2} = \frac{A_{10}}{B_{10}}\mbox{\ ,}
\end{equation}   
along with equation (7), then equation (8) can be written as 

\begin{equation}
I_{\nu_{10}} < \frac{A_{10}}{B_{10}}\left(\frac{B_{01}n_{0}}{B_{10}n_{1}}-1
\right)^{-1}\mbox{\ .}
\end{equation}  
For HD, the Einstein coefficient for spontaneous emission is
$A_{10}\simeq 5\times 10^{-8}$~s$^{-1}$ (e.g. Nakamura \& Umemura 2002).
This, in turn, implies that

\begin{equation}
n_{0} B_{01} I_{\nu_{10}} <  n_{1}A_{10} + n_{1}B_{10}I_{\nu_{10}} \mbox{\ .}
\end{equation} 

Thus, more energy is emitted into the radiation field than is absorbed from the radiation field.  If we define the energy density of the gas according to 

\begin{equation}
u_{\rmn gas} = \frac{3}{2} n k_{\rmn B} T_{\rmn gas} \mbox{\ ,}
\end{equation}
where $n$ is the total number density of the gas particles, including all species, then equation (11) implies that, with no change in the density of the gas, 

\begin{equation}
h\nu_{10}[n_{0}B_{01}I_{\nu_{10}} -  n_{1}A_{10} - n_{1}B_{10}I_{\nu_{10}}] = \frac{3}{2}nk_{\rmn B}\frac{dT_{\rmn gas}}{dt} \mbox{\ .}
\end{equation}

Despite our assumption of LTE level populations, no collisional excitation
and de-excitation terms appear in equation (13), because it expresses
the exchange of energy between two physically distinct heat reservoirs,
the CMB and our parcel of gas. In LTE, collisional excitations are exactly
balanced by de-excitations, and there would be no net heat transfer between
the two reservoirs due to these terms.
Next, we take it that the ratio of the number density of HD molecules $n_{\rmn HD}$ to the total number density of particles $n$ in the gas is given by the constant factor

\begin{equation}
X_{\rmn HD} \equiv \frac{n_{\rmn HD}}{n} \simeq \frac{n_{0} + n_{1}}{n} 
\simeq \frac{n_{0}}{n} \mbox{\ .}
\end{equation}        
We note that, since $T_{\rmn CMB} \la T_{\rmn gas} < 130$ K, 

\begin{equation}
e^{- \frac{h \nu_{10}}{k_B T_{\rmn CMB}}} \la e^{- \frac{h \nu_{10}}{k_{\rmn B} T_{\rmn gas}}} < 1 \mbox{\ .}
\end{equation}  
This implies that
 
\begin{equation}
I_{\nu_{10}} \simeq \frac{2h\nu_{10}^{3}}{c^2}e^{- \frac{h \nu_{10}}{k_{\rmn B} T_{\rmn CMB}}} = \frac{A_{10}}{B_{10}}e^{- \frac{h \nu_{10}}{k_{\rmn B} T_{\rmn CMB}}}\mbox{\ .}
\end{equation} 
For simplicity, we further assume that stimulated emission can be neglected.
The thermal evolution of the gas is then approximately described by

\begin{equation}
\frac{dT_{\rmn gas}}{dt} \simeq \frac{2h\nu_{10}A_{10}X_{\rmn HD}}{k_{\rmn B}}\left(e^{-\frac{h\nu_{10}}{k_{\rmn B} T_{\rmn CMB}}}-e^{-\frac{h\nu_{10}}{k_{\rmn B}T_{\rmn gas}}}\right)  \mbox{\ .}
\end{equation}  

It is clear from this result that if $T_{\rmn CMB} \la T_{\rmn gas} < 130$~K, with the gas cooling only by radiative decay of the excited rotational level $J$=1 to $J$=0, the temperature of the gas will asymptotically approach $T_{\rmn CMB}$.  Thus, equation (17) describes the fact that the CMB temperature is indeed a lower limit on the temperature to which a gas can cool via line emission only.  

We estimate the timescale for reaching the CMB temperature floor
to be 
\begin{eqnarray}
t_{\rmn CMB} & \simeq & \frac{1}{2 A_{10}X_{\rmn HD}}\left(
\frac{k_{\rmn B}T_{\rmn CMB}}{h\nu_{10}}\right)^2
\exp\left(\frac{h\nu_{10}}{k_{\rmn B}T_{\rmn CMB}}\right) \nonumber \\
             & \simeq &  \left(A_{10}X_{\rmn HD}\right)^{-1}\mbox{\ .} 
\end{eqnarray}
Defining a critical HD abundance by demanding $t_{\rmn CMB}\sim
t_{\rmn H}$, with $t_{\rmn H}$ being the Hubble time, we find
\begin{equation}
X_{\rmn HD, crit}\sim 10^{-8}\mbox{\ .}
\end{equation}
This estimate, derived for the limiting case of LTE level populations,
is almost independent of redshift for $z\ga 10$.
Therefore, if HD has a fractional abundance lower than $X_{\rmn HD,crit}$ $\sim 10^{-8}$, the gas will not cool to $T_{\rmn gas}  \simeq T_{\rmn CMB}$ within a Hubble time at redshifts $z \ga 10$.  Thus, a mechanism for producing large fractional abundances of HD is necessary in order for the primordial gas to cool to $T_{\rmn CMB}$ before fragmentation occurs.

\section {Evolution of primordial gas}
\subsection {First generation SN blast waves}
The passage of a SN shock through the primordial gas results in the almost complete ionization of the hydrogen for shock velocities $\ga$ 100 km s$^{-1}$, early in the radiative phase of the shocks (see Kang \& Shapiro 1992).  Fig.~3 shows the thermal evolution of primordial gas compressed, heated, and ionized by shocks with velocities in the range expected for the first radiative supernova shocks, 100 and 200 km s$^{-1}$, and at redshifts of $z$ = 10 and $z$ = 20.  The initial post-shock temperature of the gas is defined by 

\begin{equation}
T_{\rmn ps} = \frac{m_{\rmn H}u_{\rmn sh}^{2}}{3k_{\rmn B}} \mbox{\ ,}
\end{equation}
where $u_{\rmn sh}$ is the velocity of the shock, and $m_{\rmn H}$ is the mass
of a hydrogen atom.
The post-shock gas initially cools by bremsstrahlung, and then by He line emission, once the temperature has dropped enough for helium atoms to recombine.  Once hydrogen begins to recombine the temperature of the gas can drop to $\sim$10$^4$ K by atomic hydrogen line emission.  

\begin{figure}
\vspace{2pt}
\epsfig{file=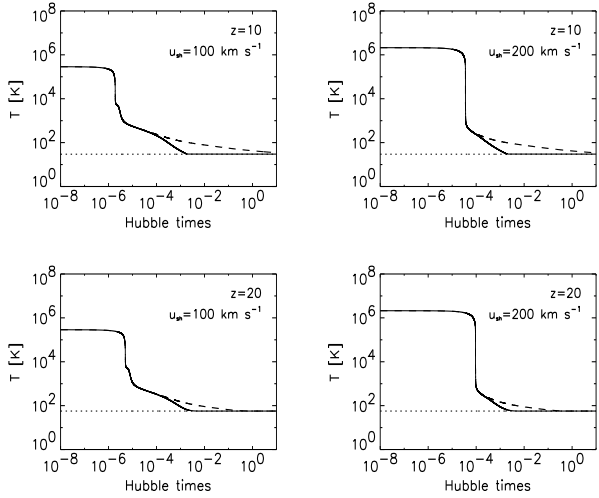,width=8.5cm,height=7.cm}
\caption{The thermal evolution of SN-shocked primordial gas.  The dashed lines show the temperature as a function of time for primordial gas without deuterium, while the solid lines show the gas temperature when deuterium is included.  The combination of redshift $z$ and shock velocity $u_{\rmn sh}$ is varied in each panel.  In each case the temperature of the gas reaches the temperature of the CMB, denoted by the dotted line, in a small fraction of the Hubble time. Each case was calculated assuming a preshock density of $n_{0}$ = 10$^2$ cm$^{-3}$.  
}
\end{figure}
For the cases shown in Fig.~3, large fractions of H$_2$ and HD form out-of-equilibrium as the gas cools below $\sim$10$^4$ K, allowing the gas to efficiently cool to temperatures $T_{\rmn gas} \la 200$ K.  In each case shown in Fig.~3, the gas containing deuterium is then able to cool to a temperature $T_{\rmn gas}$ $\simeq$ $T_{\rmn CMB}$ in $\la$ 10$^7$ yr, a fraction of the age of the Universe at redshifts $z$ $\ga$ 10.  However, without deuterium, cooling only takes place by H$_{2}$ line emission at low temperatures ($\la$ 200 K), and so it takes at least $\sim$10$^{2}$ times longer for the temperature of the deuterium-free gas to approach the CMB temperature.  Thus, while HD can rapidly cool the primordial gas to the CMB temperature, H$_2$ does not efficiently cool below $\la$ 200K.   

Although the thermal evolution in each of the cases plotted in Fig.~3 is calculated assuming a pre-shock density of $n_{0}$ = 10$^{2}$ cm$^{-3}$, we have found that the gas containing deuterium is efficiently cooled to the CMB limit over a range of pre-shock densities.  After the passage of a shock with a velocity of 200 km s$^{-1}$ at $z$ = 20, the gas cools to $T_{\rm gas} \simeq T_{\rm CMB}$ in $\sim10^{-1} t_{\rmn H}$ for $n_{\rm ps}$ = 10$^{-2}$ cm$^{-3}$ and in $\sim 10^{-2} t_{\rmn H}$ for $n_{0}$ = 1 cm$^{-3}$.  Thus, for SN shocks propagating through primordial gas in a variety of situations, sufficient HD can form to allow cooling to the temperature of the CMB in a fraction of the age of the Universe at redshifts $z$ $\ga$ 10.  As shown in Fig.~2, the fractional abundance of HD at temperatures $\la$ 200 K is $X_{\rm HD}$ $\ga$ 10$^{-6}$, well above the minimum of $X_{\rm HD,crit}$ $\sim$ 10$^{-8}$ required for the effective cooling of the primordial gas by HD.

\subsection {High-$z$ structure formation shocks}
We here distinguish between the evolution of shocks associated with SNe and with the assembly of DM halos during hierarchical structure formation. For the case of the cooling of primordial gas shock-heated during DM halo assembly, the virial velocities can be up to an order of magnitude lower than for SN shock velocities, while the resulting shock can still partially ionize the primordial hydrogen (see Shapiro \& Kang 1987). 
 
\begin{figure}
\vspace{2pt}
\epsfig{file=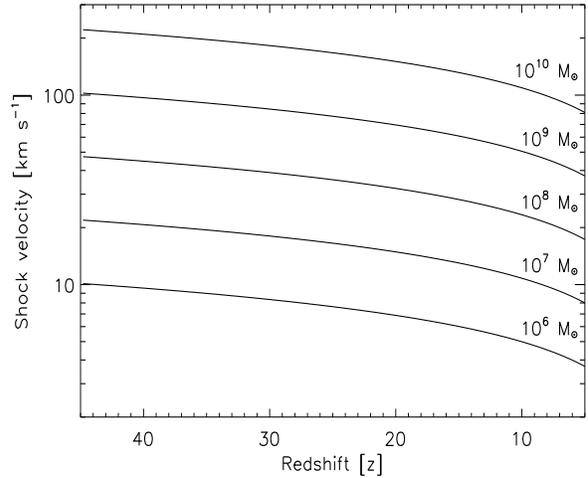,width=8.5cm,height=7.cm}
\caption{The virial velocities associated with the assembly of DM haloes during hierarchical structure formation. Each of the curves corresponds to a different halo mass. Because the densities of the virialized haloes are higher at higher redshifts, the virial velocities increase with redshift.          
}
\end{figure}
Fig.~4 shows the 
virial velocities associated with the DM halo assembly for various halo masses. Using these velocities and expressing the masses of the resultant DM haloes in terms of the $\sigma$ fluctuations they represent at a given redshift (e.g. Barkana \& Loeb 2001), Fig.~5 shows the thermal evolution of primordial gas shocked in the formation of haloes with masses between $\sim$ 10$^8$ and $\sim$ 10$^{10}$ M$_{\odot}$ at redshifts $z$ = 10 and $z$ = 15.  We again assume that the gas is strongly-shocked, and so that the post-shock evolution is isobaric.  The initial post-shock temperature is given by equation (20), and the pre-shock density of the gas is taken to be the density of baryons in DM haloes at the point of virialization, given by (e.g. Clarke \& Bromm 2003)

\begin{equation}
n_{0} \simeq 0.3  {\rm cm}^{-3}\left(\frac{1+z}{21}\right)^{3}\mbox{\ .}\end{equation}
If there are clumps of higher densities within the haloes, however, this estimate of the density will likely be too low.  Then, because the gas can cool faster at higher densities, due to the more rapid formation of H$_2$ and HD molecules, the cooling times shown in Fig.~5 may be overestimates for high density clumps within haloes.  

The initial post-shock fractional ionization of the various chemical species composing the gas were taken from Shapiro $\&$ Kang (1987) and Kang $\&$ Shapiro (1992), for the cases of shock velocities of 20, 30, 50, and 100 km s$^{-1}$.  As deuterium was not treated in this earlier work, we take it that the fractional abundances of D, D$^+$, and HD are reduced from those of H, H$^+$, and H$_2$, respectively, by a factor of the cosmic abundance of deuterium, taken to be $\sim$ 4 $\times$ 10$^{-5}$.

\begin{figure}
\vspace{2pt}
\epsfig{file=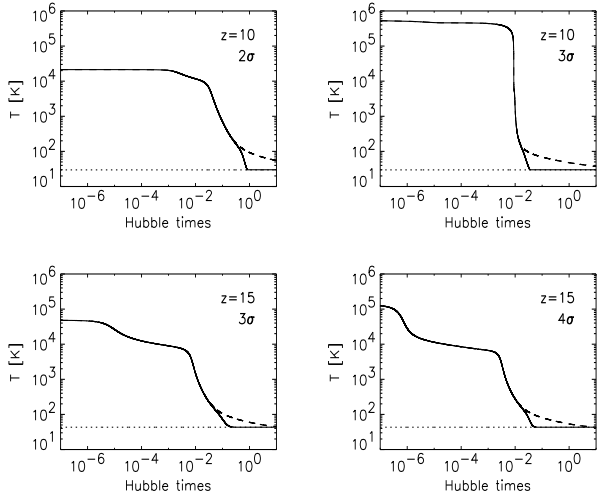,width=8.5cm,height=7.cm}
\caption{The thermal evolution of primordial gas shocked through dark matter halo collapse.  The dashed lines show the temperature as a function of time for primordial gas without deuterium, while the solid lines show the temperature of the gas when deuterium is included in the calculation.  The combination of redshift $z$ and $\sigma$ fluctuation of the DM halo is different in each panel.  In each case the temperature of the gas reaches the temperature of the CMB, denoted by the dotted line, though generally not as quickly as is the case with the higher velocity supernova shocks. The initial preshock density at a given redshift  was taken to be the virial density of baryons at that redshift. 
}
\end{figure}

As was demonstrated by Shapiro $\&$ Kang (1987), even at shock velocities $\la$ 100 km s$^{-1}$, which characterize the collapse of DM haloes at redshifts $z$ $\ga$ 10, enough out-of-equilibrium formation of molecular hydrogen takes place that the gas can cool to $\la$ 100 K within the age of the Universe, at high enough redshift (see also Murray $\&$ Lin 1990).  That H$_2$ efficiently cools the gas behind these shocks is also shown in Fig.~5, where the temperature of the primordial gas does indeed fall to $\sim$ 100 K within a Hubble time. The abundance of HD at low temperatures ($\la$ 100 K) for the case of shocks due to DM halo collapse is found to be of the same order as that found for the SN case, that is $X_{\rm HD}$ $\ga$ 10$^{-6}$, again well above the minimum necessary for efficient HD cooling.

\subsection {Unshocked primordial gas within minihaloes}
Minihaloes, with masses of the order of $10^6  {\rmn M}_{\odot}$, merge at velocities too low to strongly shock or ionize the primordial gas they contain, as can be seen from Fig.~4.  Haloes of such low mass collide at velocities of only a few kilometers per second, which are not high enough to ionize the hydrogen inside the haloes, and so are not high enough to allow significant free electron-catalyzed formation of molecules to take place.  Thus, the gas in minihaloes is expected to remain largely un-ionized and with only the primordial fractions of H$_2$ and HD until the formation of the first stars within them.  Without the large fractions of HD formed in shocked, ionized primordial gas, the gas from which the first stars formed in minihaloes at redshifts of $z\ga 20$ could not cool below $\sim 200$ K, and thus the first stars were likely very massive (Bromm et al. 1999, 2002).

\begin{figure}
\vspace{2pt}
\epsfig{file=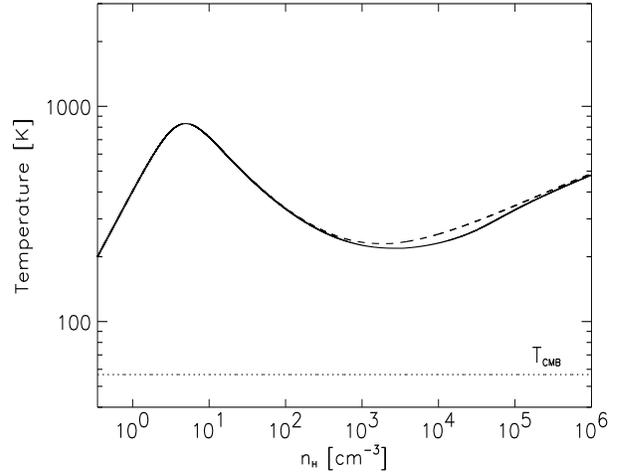,width=8.5cm,height=7.cm}
\caption{The thermal evolution of an unshocked, collapsing primordial gas cloud, such as would have resided in the first minihaloes.  Initially, the gas collapses adiabatically, starting with the density of baryons in a virialized dark matter halo at a redshift of $z$ = 20.  Upon reaching a density of $\sim$ 10 cm$^{-3}$, the gas cools by molecular rotational line-emission.  The dashed line shows the temperature evolution of the primordial gas when deuterium is not included, leaving the gas to cool only by H$_{2}$ line-cooling, which is effective at cooling the gas only down to $\sim$ 200 K.  The solid line shows the temperature evolution of the gas when deuterium is included in the calculation.  Although, in principle, HD can efficiently cool the gas to the temperature of the CMB, denoted by the dotted line, unshocked gas of primordial composition does not contain sufficient HD ($X_{\rmn HD}$ $\sim$ 10$^{-9}$) for it to be an efficient coolant.    
}
\end{figure}

Fig.~6 shows the thermal evolution of unshocked gas of primordial composition with an initial ionization fraction of $x_{\rmn e}$ = 10$^{-4}$ which collapses under its own gravity, its density evolving according to $dn/dt = n/t_{\rmn ff}$, where $n$ is the number density of the gas and $t_{\rmn ff}$ is the free-fall time, given by 

\begin{equation}
t_{\rmn ff} = \left(\frac{3\pi}{32G\rho}\right)^{1/2} \mbox{\ ,}
\end{equation} 
where $\rho = m_{\rmn H}n$.
Here, the initial abundances for this calculation were taken to be the primordial abundances of the standard cosmological model given by Galli \& Palla (1998), at redshifts of $z \sim 20$.  We take the initial temperature of the gas to be 200 K (e.g. Bromm et al. 2002), and the initial density of the gas is taken to be the density of baryons within DM haloes at the point of virialization, given by equation (21).

The gas at first is heated by adiabatic compression as it collapses.  Once reaching a density of $\sim 10$~cm$^{-3}$ and a temperature of $\sim 10^3$ K, the gas begins to cool via H$_2$ line emission, and the temperature drops again to $\sim 200$ K.  As Fig.~6 shows, the fraction of HD stays too low for HD to efficiently cool the gas below these temperatures.  We find that for almost a Hubble time after the collapse has begun, the abundance of HD has a value of $X_{\rmn HD} \la 10^{-8}$, which is below the minimum abundance necessary for effective HD cooling.  Because the gas in this case is never ionized, the reactions (1)-(3), through which HD forms in the case of strongly-shocked primordial gas, are suppressed and HD is not effectively produced.  Thus, cooling proceeds largely only through H$_2$ line emission and the gas temperature never drops below $\sim$ 200 K.

\subsection {HD cooling in relic H II regions}
The first stars which formed in minihaloes were likely strong sources of ionizing radiation due to their high mass (e.g. Bromm \& Larson 2004).  The possibility of a second-generation of stars forming in the so-called relic H II regions, carved out by the copious amounts of ionizing radiation emitted by first generation massive stars, has recently received some attention.  In particular, it has been argued that the ionized conditions in relic H II regions allow for out-of-equilibrium production of large fractions of H$_2$ (O'Shea et al. 2005).  Given that out-of-equilibrium fractions of HD can be created from H$_2$ through reactions (1)-(3), large fractional abundances of HD may be created in relic H II regions, as well.  Thus, HD may be an important coolant of the primordial gas in such regions.    
\begin{figure}
\vspace{2pt}
\epsfig{file=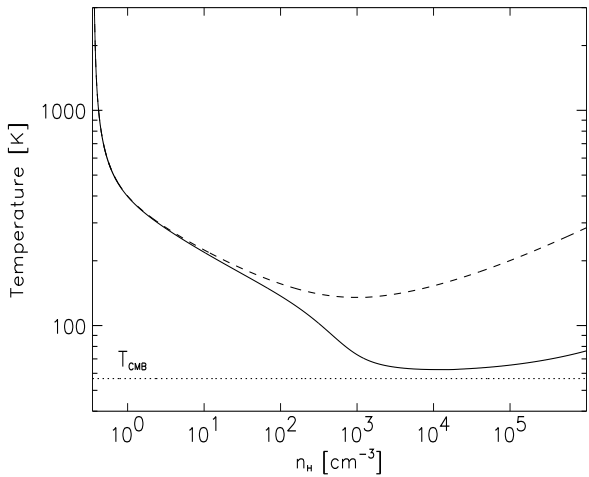,width=8.5cm,height=7.cm}
\caption{The thermal evolution of a primordial relic H II region, in which an initially ionized primordial gas cloud collapses under its own gravity.  The initial temperature of the gas was set to $T_{\rmn gas} = 3 \times 10^{4}$~K, typical for an H~II region surrounding a massive Pop~III star at $z \simeq 20$.  The dashed line denotes the temperature of the gas as a function of its increasing density, neglecting HD.  The solid line shows the temperature evolution of the gas with HD cooling included.  The free electrons, left to recombine following the extinguishing of the ionizing radiation from the hot star, catalyze the formation of H$_{2}$, and consequently of HD.  The abundance of HD becomes high enough to cool the gas to close to the CMB temperature at $z$ = 20.          
}
\end{figure}
Fig.~7 shows the thermal evolution of primordial gas in a relic H II region.  The initial temperature of the gas is taken to be $3 \times 10^{4}$~K, a temperature typical of an H II region surrounding a massive Pop~III star (e.g. Storey \& Hummer 1995).  We thus take the hydrogen in the gas to be initially fully ionized and the fractional abundances of H$_2$ and HD to be initially zero.  The initial density, as in the cases of gas shocked in DM halo assembly and gas within minihaloes, is taken to be the density of baryons within DM haloes at the point of virialization, given by equation (21), for a redshift of $z \sim 20$.  This density is consistent with the density of $\la$ 1 cm$^{-3}$ expected to result from the dynamical expansion of the first H II regions due to photoheating from a central massive star (see Kitayama et al. 2004; Alvarez et al. 2005).  As in the case of unshocked gas within minihaloes, the evolution of the density $n$ of the gas is described by $dn/dt = n/t_{\rm ff}$.

The effect of the out-of-equilibrium free electron-catalyzed production of H$_2$ is evident, as the temperature of the primordial gas lacking deuterium falls to slightly lower values than in the case of unshocked, un-ionized primordial gas shown in Fig.~6.  The larger fraction of H$_2$ in the relic H II region case results in more effective H$_2$ cooling.  In turn, a large out-of-equilibrium abundance of HD results, as well, with the fractional abundance of HD exceeding $X_{\rmn HD} \sim 10^{-5}$ at the lowest temperatures ($\la 100$~K).  With this enhanced fraction of HD the gas is able to cool almost to the temperature of the CMB at $z$ $\simeq$ 20, despite the adiabatic compression that heats the gas at high densities.       

Fig.~8 summarizes the evolution of the fractional abundance of HD, $X_{\rmn HD}$, for the cases previously described.  HD cooling is significant whenever  
$X_{\rmn HD}> X_{\rmn HD, crit}$ is reached in a fraction of the Hubble time. 

\begin{figure}
\vspace{2pt}
\epsfig{file=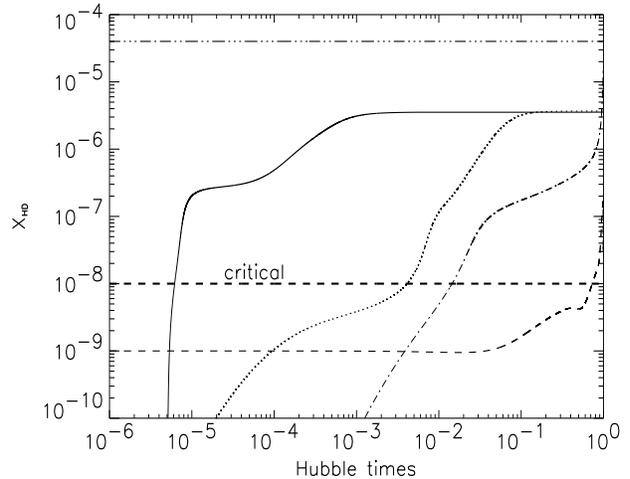,width=8.5cm,height=7.cm}

\caption{The evolution of the abundance of HD, $X_{\rm HD}$, in primordial gas which cools in four distinct situations.  The solid line corresponds to gas with an initial density of 100 cm$^{-3}$ which is compressed and heated by a supernova shock with velocity  $u_{\rm sh}$ = 100 km s$^{-1}$ at $z$ = 20.  The dotted line corresponds to gas at an initial density of 0.1 cm$^{-3}$ shocked in the formation of a 3$\sigma$ fluctuation dark matter halo at $z$ = 15.  The dashed line corresponds to unshocked, un-ionized primordial gas with an initial density of 0.3 cm$^{-3}$ collapsing inside a minihalo at $z$ = 20.  Finally, the dash-dotted line shows the fraction of HD in primordial gas collapsing from an initial density of 0.3 cm$^{-3}$ inside a relic H II region at $z$ = 20.   The temperature evolution in these situations is illustrated in figures 3, 5, 6, and 7, respectively.  The horizontal line at the top denotes the cosmic abundance of deuterium.  The critical abundance of HD, $X_{\rmn HD,crit}$, denoted by the bold dashed line, is that above which primordial gas can cool to the CMB temperature within a Hubble time, as given by equation (19).  Although the HD abundance in the relic H II region does exceed $X_{\rmn HD,crit}$, adiabatic heating prevents the temperature of the gas in the relic H II region from cooling to the CMB temperature.  The supernova-shocked gas and the gas shocked in the merging of DM haloes evolve isobarically and so adiabatic heating has much less effect on the temperature evolution in these situations. 
}
\end{figure}

\section {Fragmentation of primordial gas}
\subsection{Characteristic stellar mass for SN-shocked gas}
The characteristic mass of Pop~II.5 stars formed from SN-shocked primordial gas can now be estimated, assuming that the shock is strong, so that the shocked gas evolves isobarically, and that the gas cools to the temperature of the CMB.  Following Mackey et al. (2003), we assume that the immediate progenitor of a star is a self-gravitating gas cloud with a mass well-approximated by the Bonnor-Ebert mass, given as (e.g. Palla 2002)

\begin{equation}
M_{\rmn BE}\simeq 700 {\rmn M}_{\odot} \left(\frac{T_{\rmn final}}{200 {\rmn \,K}}
\right)^{3/2}\left(\frac{n_{\rmn final}}{10^{4}{\rmn cm}^{-3}}\right)^{-1/2} \mbox{\ ,}
\end{equation} 
where $n_{\rmn final}$ is the final density and $T_{\rmn final}$ is the final temperature of the post-shock gas prior to fragmentation.
Isobaric density evolution gives

\begin{equation}
n_{\rmn final} \simeq \frac{T_{\rmn ps}n_{0}}{T_{\rmn final}} \mbox{\ ,} 
\end{equation}
where $n_{0}$ is the pre-shock density of the gas and $T_{\rmn ps}$ is the initial post-shock gas temperature.
This initial post-shock temperature is roughly

\begin{equation}
T_{\rmn ps} \simeq \frac{m_{\rmn H}u_{\rmn sh}^{2}}{k_{\rmn B}} \mbox{\ ,}
\end{equation}
where $k_{\rmn B}$ is the Boltzmann constant, $u_{\rmn sh}$ is the shock velocity, and $m_{\rmn H}$ is the mass of the hydrogen atom. 
The CMB floor defines the final temperature as follows:

\begin{equation}
T_{\rmn final} = T_{\rmn CMB} = \mbox{2.7~K}(1+z)\mbox{\ .}
\end{equation}
Substituting equations (24), (25), and (26) into equation (23) gives

\begin{equation}
M_{\rmn BE} \simeq 4 {\rmn M}_{\odot} \left(\frac{1+z}{21}\right)^{2}\left(\frac{n_{0}}{10^{2}{\rmn cm}^{-3}}\right)^{-1/2} \left(\frac{u_{\rmn sh}}{200{\rmn km} {\rmn s}^{-1}}\right)^{-1} \mbox{\ .} 
\end{equation}
Next, as an estimate of a typical preshock density of baryonic material in a DM halo, assume the density at the point of virialization of the halo, given in Clarke \& Bromm (2003) as 

\begin{equation}
n_{0} \simeq 0.3  {\rmn \ cm}^{-3}\left(\frac{1+z}{21}\right)^{3}\mbox{\ .}
\end{equation}

Define the characteristic Pop~II.5 stellar mass as
$M_{\rmn char} \simeq \alpha M_{\rmn BE}$, with the parameter $\alpha = 0.3$ describing the efficiency with which the mass in the cloud is incorporated into stars.
Thus, M$_{\rmn BE}$ gives a robust upper limit on the masses of Pop~II.5 stars.  
This value for $\alpha$ is close to that inferred for the formation of stars in the
present-day Universe, although it is in general a function of the specific accretion physics involved in star formation and is not yet known with any certainty for the case of primordial star formation (e.g. McKee \& Tan 2002; Bromm \& Loeb 2004).  
Then, using equations (27) and (28), we find

\begin{equation}
M_{\rmn char} \simeq 20  {\rmn M}_{\odot} \left(\frac{1+z}{21}\right)^{1/2} \left(\frac{u_{\rmn sh}}{200{\rmn km} {\rmn s}^{-1}}\right)^{-1} \mbox{\ .} 
\end{equation}
This formula for the characteristic Pop~II.5 stellar mass is the equivalent to equation (9) in Mackey et al. (2003), which was calculated neglecting the effects of HD cooling on the shocked primordial gas.  Because HD cooling allows shocked primordial gas to efficiently cool to the temperature of the CMB, the characteristic mass of Pop~II.5 stars is {\it redshift dependent}, and strong shocks in the primordial gas will thus yield higher mass stars, on average, the earlier in the history of structure formation they occur.  As well, higher velocity radiative shocks will tend to yield lower mass Pop~II.5 stars.  Fig.~9 shows the characteristic Pop~II.5 stellar mass as a function of redshift for various SN shock velocities.   

\begin{figure}
\vspace{2pt}
\epsfig{file=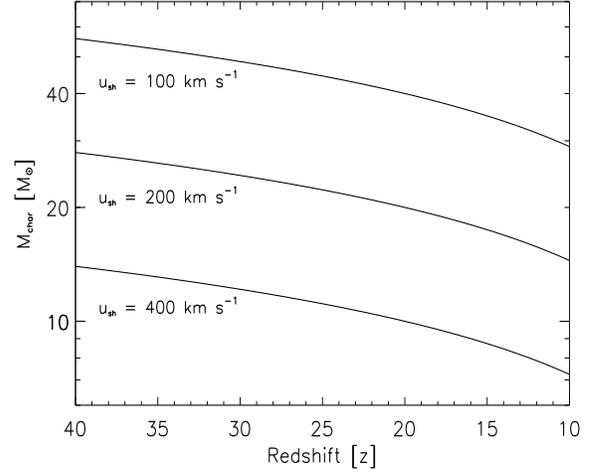,width=8.5cm,height=7.cm}
\caption{The characteristic mass of Pop~II.5 stars, formed from primordial gas shock-compressed by the first supernovae.  As shown in equation (29), this characteristic mass varies inversely with the velocity of the supernova shock $u_{\rmn sh}$, while increasing with redshift.  That the masses of these stars explicitly depend on redshift is due to the fact that HD cools the shocked primordial gas to the temperature of the CMB.  For each of the shock velocities shown here, the density of the preshock gas was taken to be the virial value given in equ. (21).
}
\end{figure}

\subsection{Characteristic stellar mass for DM halo collapse}
To calculate the characteristic mass of stars formed in DM halo assembly, we assume that the shocked gas, with a post-shock temperature $T_{\rmn ps}$ roughly equal to the virial temperature of the halo, cools isobarically, following the treatment of Clarke \& Bromm (2003).  The pressure of the gas as it cools, defined as that necessary to balance the weight of the overlying baryons, is given by these authors as

\begin{equation}
P_{\rmn vir} \sim 10^7 k_{\rmn B} M_{8}^{2/3} \left(\frac{1+z}{21}\right)^4 {\rmn K~cm}^{-3} \mbox{\ ,} 
\end{equation}       
where $M_{x}=M/10^{x}{\rmn M}_{\odot}$. 
In taking this value for the pressure, we have assumed that the mass of the halo is higher than the critical mass above which centrifugal effects become important in slowing the collapse of the gas in the halo, given by Clarke \& Bromm (2003) as 

\begin{equation}
M_{\rmn crit} \simeq 10^{8} {\rmn M}_{\odot} \left(\frac{1+z}{21}\right)^{3/2} \lambda_{0
.1}^{-3}\mbox{\ .}
\end{equation}
We here assume that the spin parameter has the fiducial value of $\lambda_{0.1}$ = 1.  Again, we have that $T_{\rmn final}$ = $T_{\rmn CMB}$ = 2.7$~{\rmn K}$ (1+$z$).  Thus, the density of the gas when the temperature has dropped to $T_{\rm final}$ is given by 

\begin{equation}
n_{\rmn final} \simeq P_{\rmn vir}/(k_{\rmn B}T_{\rmn CMB}) \mbox{\ .} 
\end{equation}  
Using these values for the final temperature and density in equation (23), we obtain the characteristic stellar mass for the case of stars formed in collapsing DM haloes as

\begin{equation}
M_{\rmn char} \simeq 10 M_{8}^{-1/3}{\rmn M_{\odot}}  \mbox{\ ,} 
\end{equation}
where we have again used $M_{\rmn char} \simeq \alpha M_{\rmn BE}$, with $\alpha = 0.3$. The characteristic mass of stars formed in the assembly of DM haloes with masses $\ga 10^8$ M$_{\odot}$ depends only on the halo mass.  Thus, stars formed in this process are expected to have characteristic masses of the order of 10 M$_{\odot}$, comparable to the SN-shocked case.

\section {Summary and Conclusions}
We have investigated the importance of HD as a coolant of the primordial gas in four distinct cases:  the evolution of supernova-shocked gas, with shock velocities $\ga$ 100 km s$^{-1}$;  the evolution of gas shocked during DM halo collapse, with shock velocities $\la$ 100 km s$^{-1}$;  freely collapsing unshocked, un-ionized gas clouds within minihaloes; and freely collapsing clouds of ionized gas within relic H II regions.  In the cases where the primordial gas is ionized, HD is an important coolant.  Thus, behind strong shocks and in relic H II regions, the primordial gas is able to cool by HD line emission to temperatures $\la$ 100 K.  In the case of strong shocks, HD cooling can lower the temperature of the primordial gas to the temperature of the CMB in a fraction of the age of the Universe at redshifts $z$ $\ga$ 10.  In much of the literature, it is assumed that metals are required to lower the temperature of interstellar gas to the CMB temperature (e.g. Bromm et al. 2001; Schneider et al. 2002; Clarke \& Bromm 2003).  We emphasize that through HD cooling {\it  gas of purely primordial composition, without any metals, can cool to the CMB temperature}, the lowest temperature attainable by radiative cooling.  All that is required is for a large enough abundance of HD to be produced, which we have shown can be done under strongly-shocked conditions in which the primordial gas is ionized and evolves isobarically.     

We have calculated the characteristic mass of primordial stars formed from gas that has cooled to the CMB temperature floor. We estimate that Pop~II.5 stars, formed from gas shocked either through SN explosions or during the assembly of the first dwarf galaxies, have characteristic masses that are redshift dependent and of the order of 10 M$_{\odot}$.

The implications of this ability of the primordial gas to effectively cool to low temperatures are numerous.  HD cooling offers a mechanism by which low mass metal-poor stars can be formed very early in the history of the Universe, at redshifts $z$ $\ga$ 10.  Thus, HD cooling may play a role in the formation of the most metal-poor stars that have been observed.  In particular, it has been suggested that the metal abundance patterns of two recently discovered extremely metal poor halo stars, HE0107-5240 and HE1327-2326, can be explained by these stars having formed from gas enriched by the elements released in the type II supernova explosion of a 20-25 M$_{\odot}$ star (Christlieb et al. 2002; Frebel et al. 2005).  While the first generation of stars which formed in minihaloes of mass $10^6$ M$_{\odot}$ would likely have been far more massive than this ($\ga 100$ M$_{\odot}$), the characteristic mass that we calculate for Pop~II.5 stars falls just in this mass range.  Therefore, it may have been the supernova explosions of Pop~II.5 stars that dispersed metals into the primordial gas that later formed the most metal poor stars that are observed today.

\begin{figure}
\vspace{2pt}
\epsfig{file=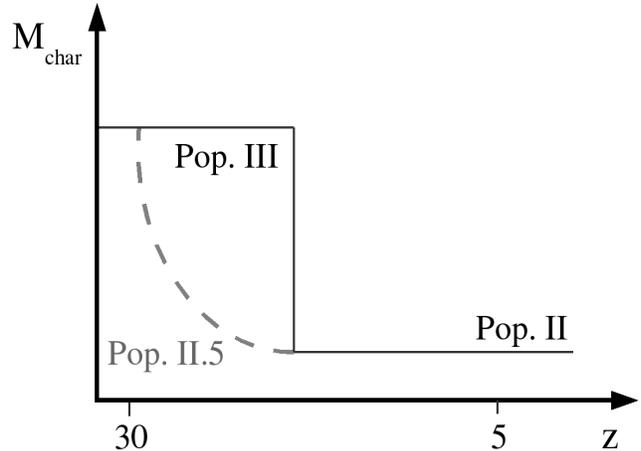,width=8.5cm,height=6.cm}
\caption{The characteristic mass of stars as a function of redshift.  Pop~III stars, formed from unshocked, un-ionized primordial gas are characterized by masses of the order of 100 M$_{\odot}$.  Pop~II stars, formed by gas which is enriched with metals, are formed at lower redshifts and have characteristic masses of the order of 1 M$_{\odot}$.  Pop~II.5 stars, formed from shocked primordial gas, have characteristic masses reflecting the fact that they form from gas that has cooled to the temperature of the CMB.  Thus, the characteristic mass of Pop~II.5 stars, given by equation (29), is a function of redshift and is of the order of 10 M$_{\odot}$.  }
\end{figure}
As we have shown, the evolution of the primordial gas following the DM halo collapse that formed the first dwarf galaxies may have allowed for the formation of primordial stars with masses of the order of 10 M$_{\odot}$.  The formation of the oldest globular clusters, which formed within $\sim$ 10$^9$ yr after the big bang, must have required the presence of metal coolants to account for the observed abundance of low-mass stars in them (see Ashman \& Zepf 1992;  Kang et al. 1990; Bromm \& Clarke 2002).

Overall, the fact that primordial gas can cool to the CMB temperature by HD line emission suggests that primordial stars with masses at least an order of magnitude lower than those characteristic of the first stars could have formed.  If this is indeed the case, then we obtain a redshift dependence for the characteristic stellar mass of three distinct modes of star formation, shown schematically in Fig.~10.  Pop~III stars, formed in the first minihaloes out of un-ionized gas that could effectively cool only to $\sim$ 200 K through H$_2$ line emission, would have been the first stars, with masses of the order of $\sim$ 100 M$_{\odot}$.  Then, as these first stars ended their short lives by exploding as supernovae, the strong shocks thus generated would have initiated the formation of large fractions of HD, allowing the primordial gas to cool to the CMB temperature floor, and so making possible the formation of Pop~II.5 stars with masses of the order of 10 M$_{\odot}$.  As well, shocks associated with the formation of primordial dwarf galaxies, with masses $\ga$ 10$^8$ M$_{\odot}$, could have triggered HD cooling, as well, yielding stars with masses of the order of 10 M$_{\odot}$.  Finally, with the dispersion of sufficient amounts of metals into the interstellar medium, Pop~II star formation would have set in, with gas clouds effectively cooling to the CMB temperature floor at redshifts $z$ $\ga$ 3 and to $\sim$ 10 K up to the present day, the latter temperature set by dust heating (see Clarke \& Bromm 2003; Mackey et al. 2003).

In the present calculations we have neglected radiative transfer and the effects of photodissociation on chemical abundances.  Because the HD photodissociation time is much longer than the HD formation time in a LW radiation field produced solely by the shocked gas in the cases we study, we expect that it is safe to neglect the effects of photodissociating radiation when the only UV flux is from the shocked gas. 
 However, with a strong UV background, such as might be generated by first generation massive stars in a primordial dwarf galaxy, photodissociation of both HD and H$_2$ will become much more important in determining the thermal and chemical evolution of the primordial gas.  The effects of HD and H$_2$ self-shielding, as well as of HD shielding by H$_2$, will have to be carefully included in any calculation of the cooling of the gas.
Thus, more detailed future work will have to account for the effects of a possible strong UV background and of HD self-shielding, in order to better discern the process of the cooling and fragmentation of shock-compressed primordial gas.  In particular, if the primordial gas cools in an environment with a strong enough UV background, then the photodestruction of HD may be so effective as to keep $X_{\rmn HD}$, the abundance of HD, below $X_{\rmn HD,crit}$, and so to prevent the primordial gas from cooling to the the temperature of the CMB within a Hubble time.  In this case, metals would be required for gas to cool to the CMB temperature floor.  

In addition to a detailed treatment of the photodissociation and self-shielding of HD, consideration of the various possible heating mechanisms in the high-z Universe, such as cosmic rays from the first supernovae, must be included in a more thorough treatment of the cooling of the primordial gas to temperatures approaching the CMB limit (e.g. Rollinde, Vangioni \& Olive 2005).  
Our idealized treatment serves to illustrate the importance of HD for primordial star formation in varying environments, specifically, for the efficient cooling of shocked primordial gas to the temperature of the CMB.  To more fully explore the importance of the HD coolant, we will carry out three
dimensional numerical simulations of the fragmentation process in shocked
primordial gas.

\section*{Acknowledgments}
We would like to thank Avi Loeb, Paul Shapiro, Simon Glover, and Naoki Yoshida for helpful comments and discussion. This work was supported in part by NASA {\it Swift} grant NNG05GH54G.

\end{document}